\documentclass[lettersize,journal]{IEEEtran}

\usepackage{amsmath,amssymb}

\usepackage[T1]{fontenc}
\usepackage{graphicx}
\usepackage{booktabs}   %
\usepackage[table]{xcolor} %
\usepackage{adjustbox}  %
\usepackage{arydshln}   %
\usepackage[hidelinks]{hyperref}

\usepackage{lipsum}

\usepackage{amsmath,amsfonts}
\usepackage{algorithmic}
\usepackage{algorithm}
\usepackage{array}
\usepackage[caption=false,font=normalsize,labelfont=sf,textfont=sf]{subfig}
\usepackage{textcomp}
\usepackage{stfloats}
\usepackage{url}
\usepackage{verbatim}
\usepackage{graphicx}
\usepackage{cite}
\begin{document}

\title{Dynamic Feature Fusion: Combining Global Graph Structures and Local Semantics for Blockchain Fraud Detection}

\author{Zhang Sheng*,~Liangliang Song*, Yanbin Wang
\thanks{Zhang Sheng is with Hefei University of Technology (e-mail: dcszhang@foxmail.com).}%
\thanks{Liangliang Song is with Xidian University (e-mail: songliangl@stu.xidian.edu.cn).}%
\thanks{Yanbin Wang is with Xidian University (e-mail: wangyanbin15@mails.ucas.ac.cn). Yanbin Wang is the corresponding author.}%
\thanks{*These authors contributed equally to this work.}}

\maketitle

\begin{abstract}
The advent of blockchain technology has facilitated the widespread adoption of smart contracts in the financial sector. However, current fraud detection methodologies exhibit limitations in capturing both global structural patterns within transaction networks and local semantic relationships embedded in transaction data. Most existing models focus on either structural information or semantic features individually, leading to suboptimal performance in detecting complex fraud patterns.In this paper, we propose a dynamic feature fusion model that combines graph-based representation learning and semantic feature extraction for blockchain fraud detection. Specifically, we construct global graph representations to model account relationships and extract local contextual features from transaction data. A dynamic multimodal fusion mechanism is introduced to adaptively integrate these features, enabling the model to capture both structural and semantic fraud patterns effectively.
We further develop a comprehensive data processing pipeline, including graph construction, temporal feature enhancement, and text preprocessing. Experimental results on large-scale real-world blockchain datasets demonstrate that our method outperforms existing benchmarks across accuracy, F1 score, and recall metrics. This work highlights the importance of integrating structural relationships and semantic similarities for robust fraud detection and offers a scalable solution for securing blockchain systems.
\end{abstract}






\section{Introduction}
\label{introduction}
\IEEEPARstart{B}{lockchain} technology has developed rapidly in recent years and has triggered far-reaching changes in several fields, especially in the financial industry \cite{pal2021blockchain}. However, as the popularity of blockchain applications grows, so does the significant increase in fraudulent behaviors it has brought about, with serious implications for society \cite{bhowmik2021comparative}. Blockchain technology, due to its decentralization and transparency, has become a tool for unscrupulous individuals to exploit, although it provides greater security and efficiency in financial transactions \cite{wenhua2023blockchain}. For example, the application of blockchain technology in the supply chain is seen as an effective means to enhance transparency and traceability, but it also faces a crisis of social trust due to fraudulent behavior \cite{bhutta2021survey}. In addition, the increase in fraudulent and illegal activities poses new challenges to the global economy as blockchain expands and its applications grow, especially in high-risk financial transactions \cite{lai2021evaluating}. Therefore, despite its enormous potential, blockchain technology comes with social and regulatory issues that need to be addressed to ensure its safe and sustainable development \cite{article}.

As one of the most widely used blockchain platforms \cite{zheng2024dappscan}, Ethereum's fraud detection methods are constantly evolving to cope with a variety of complex frauds. Currently, commonly used methods include Graph Neural Networks (GNNs), deep learning models, and machine learning-based classification algorithms. These methods excel at handling complex transactional network relationships and identifying suspicious accounts. For example, GNNs can effectively capture structured information in transaction networks \cite{ancelotti2024review}, while deep learning models based on Long Short-Term Memory (LSTM) networks excel at processing time series patterns in transaction data \cite{roy2022exploiting}. However, current research has some limitations, especially in the following two aspects:

\begin{enumerate}
    \item \textbf{Local semantic similarity information:} In blockchain, there is usually some correlation between the transaction data of an account \cite{wu2021analysis}. For example, a phishing account may make a large number of transactions in a short period of time, where the transactions before and after an account in a certain time period have specific semantic correlations. However, it is difficult for existing methods to fully capture and utilise this local semantic similarity information, which may lead to insufficient accuracy in fraud detection.
    \item \textbf{Global transaction account network information:} Fraudulent and normal accounts often imply important global information in the transaction network graph in the blockchain \cite{elmougy2023demystifying}. Although some graph models are able to recognise these patterns, effectively combining them with local semantic similarity information remains a challenge. This lack of combination limits the performance of existing methods in complex fraud detection.
\end{enumerate}

Integrating useful information is a highly promising direction to address the above issues \cite{liu2023survey}. In this study, we propose a deep learning framework with multimodal fusion for fraud detection in blockchain transaction data. Compared with traditional methods, the proposed approach effectively captures both global structural relationships in transaction networks and local semantic patterns embedded in transaction records, achieving higher accuracy and robustness in detecting complex fraud behaviors.

Specifically, we first construct a global account interaction graph to represent the relationships between blockchain transaction accounts. Each node in the graph corresponds to an account, while the edges capture the transaction behaviors, such as frequency, transaction value, and temporal patterns. To extract meaningful structural features from this graph, we employ graph-based representation learning, which aggregates information from neighboring accounts to capture both direct and indirect relationships within the transaction network. This step enables the model to uncover global interaction patterns that are indicative of fraudulent behaviors.

Simultaneously, we process the semantic information embedded in transaction data using a pre-trained text representation model. The model converts textual descriptions, such as transaction amounts, smart contract details, and other metadata, into high-dimensional feature vectors. This process allows the model to identify local contextual relationships, such as recurring transaction patterns or anomalous textual characteristics associated with suspicious accounts.

To effectively leverage both structural and semantic insights, we propose a dynamic feature fusion mechanism that adaptively integrates these two feature spaces. The mechanism learns to balance global network structures and local transaction semantics based on their relative importance for each transaction, enabling the model to detect subtle and complex fraud patterns with high accuracy.

By combining these complementary perspectives—global structural relationships and local semantic features—our approach significantly improves the robustness and precision of fraud detection. Experimental results on real-world blockchain datasets demonstrate that the proposed ETH-GBERT model achieves state-of-the-art performance. Specifically, on the Multigraph dataset, the model achieved an F1 score of 94.71\%, significantly outperforming the best-performing baseline (Role2Vec, F1 score of 74.13\%), with an improvement of 20.58\%. On the Transaction Network dataset, ETH-GBERT achieved an F1 score of 86.16\%, representing a substantial enhancement over the next best model (Role2Vec, F1 score of 71.39\%). On the B4E dataset, the model demonstrated superior recall (89.57\%) and precision (90.84\%), far surpassing other baseline models, whose F1 scores did not exceed 74.25\%. These results highlight the model's effectiveness in capturing complex fraud patterns, its robustness in handling imbalanced data distributions, and its ability to integrate structural and semantic features dynamically. The corresponding code link is available at \href{https://github.com/dcszhang/Dynamic_Feature}{https://github.com/dcszhang/Dynamic\_Feature}

The main contributions of this study are as follows:
\begin{enumerate}
        \item A dynamic multimodal fusion model is proposed, which innovatively combines graph structure information with text semantic similarity information to enhance the fraud detection performance in blockchain smart contracts.
    \item A complete set of data processing flow is developed, including the extraction of transaction data, the generation of adjacency matrix, and the processing of text representation based on BERT, which provides a useful reference for other blockchain applications.
    \item The effectiveness of the proposed method is verified through experiments, and the results show that the method performs well in detecting complex frauds and significantly outperforms existing benchmark models.
\end{enumerate}

\section{Related Work}
In recent years, with the rapid development of blockchain technology, the frequent occurrence of fraud in blockchain networks has become a global challenge. Researchers and developers have developed a variety of fraud detection methods to address these challenges and ensure the security and reliability of blockchain systems \cite{osterrieder2024enhancing}.

\subsection{Graph-based Fraud Detection}
In blockchain networks, transaction data usually has a complex relational structure, and graph-based models can effectively capture these complex relationships and excel in fraud detection. Especially in blockchain platforms like Ethereum, Graph Neural Networks (GNNs) are widely used to detect fraud. For example, Tan \cite{tan2021graph} proposed a model based on Graph Convolutional Networks (GCNs) for detecting fraud from Ethereum transaction records. They classified addresses as legitimate or fraudulent by constructing a transaction network and extracting node features. In addition, Kanezashi \cite{kanezashi2022ethereum} investigated the application of Heterogeneous GNNs (Heterogeneous Graph Neural Networks) in Ethereum transaction networks, focusing on handling large-scale networks and the label imbalance problem. Li \cite{li2022phishing} also proposed a phishing detection framework called PDGNN, based on the Chebyshev-GCN, which can detect fraud in Ethereum transaction networks by extracting transaction subgraphs and training a classification model, effectively distinguishing normal accounts from phishing accounts in large-scale Ethereum networks. Wang \cite{wang2022tsgn} proposed the Transaction SubGraph Network (TSGN) framework to enhance phishing detection in Ethereum by constructing transaction subgraphs that capture essential features of transaction flows. Hou \cite{hou2022detecting} proposed an Ethereum phishing detection method based on GCN and Conditional Random Field (CRF). This method first utilizes DeepWalk to generate initial features for each account node in the transaction graph, then employs GCN to learn graph-structured representations, capturing the transactional relationships between accounts. To enhance classification performance, a CRF layer further encourages similar nodes to learn similar representations.

\subsection{Fraud Detection Based on Time Series Data}
Time series data analysis plays an important role in blockchain fraud detection, especially in processing transaction records and detecting abnormal behaviours. Ethereum, as one of the major blockchain platforms, contains a large amount of time-series information, such as transaction time, frequency, and Value fluctuations, which can be used to identify potential fraudulent behaviours. Hu \cite{hu2021transaction} investigated the application of time-series analysis methods based on the Long Short-Term Memory (LSTM) network in Ethereum smart contracts. Another study by Farrugia \cite{farrugia2020detection} proposed the use of the XGBoost model combined with time series features for illegal account detection in Ethereum. The study highlighted the importance of time series features, such as time intervals, in identifying illegal accounts by extracting key time series features and combining them with a machine learning model. Pan \cite{pan2024ethershield} proposed a system called EtherShield, which combines time interval analysis and contract code features to detect malicious behaviour on the Ethereum blockchain.

\subsection{Hybrid Methods}
Hybrid methods integrate various types of information, such as graph data, time-series data, and semantic information, achieving higher detection accuracy and robustness, effectively identifying complex and dynamic fraud patterns in Ethereum malicious transaction detection. Li \cite{li2022ttagn} proposed the Temporal Transaction Aggregation Graph Network (TTAGN) for Ethereum phishing detection, utilizing temporal transaction data to improve accuracy. TTAGN combines temporal edge representation, edge-to-node aggregation, and structural enhancement to capture transaction patterns and network structure, outperforming existing methods on real-world datasets. Wen \cite{wen2023novel} proposed a hybrid feature fusion model named LBPS for phishing detection on Ethereum, combining LSTM-FCN and BP neural networks. This model integrates features extracted through manual feature engineering and transaction records analysis, using BP neural networks to capture hidden relationships between features and LSTM-FCN networks to extract temporal features from transaction data. Chen \cite{chen2024ethereum} proposed the DA-HGNN model, a hybrid graph neural network with data augmentation for Ethereum phishing detection. This model utilizes data augmentation to address sample imbalance, integrates Conv1D and GRU-MHA to extract temporal features, and employs SAGEConv to capture structural features from the transaction graph.

\begin{figure*}[!ht]
    \centering
    \includegraphics[width=\linewidth]{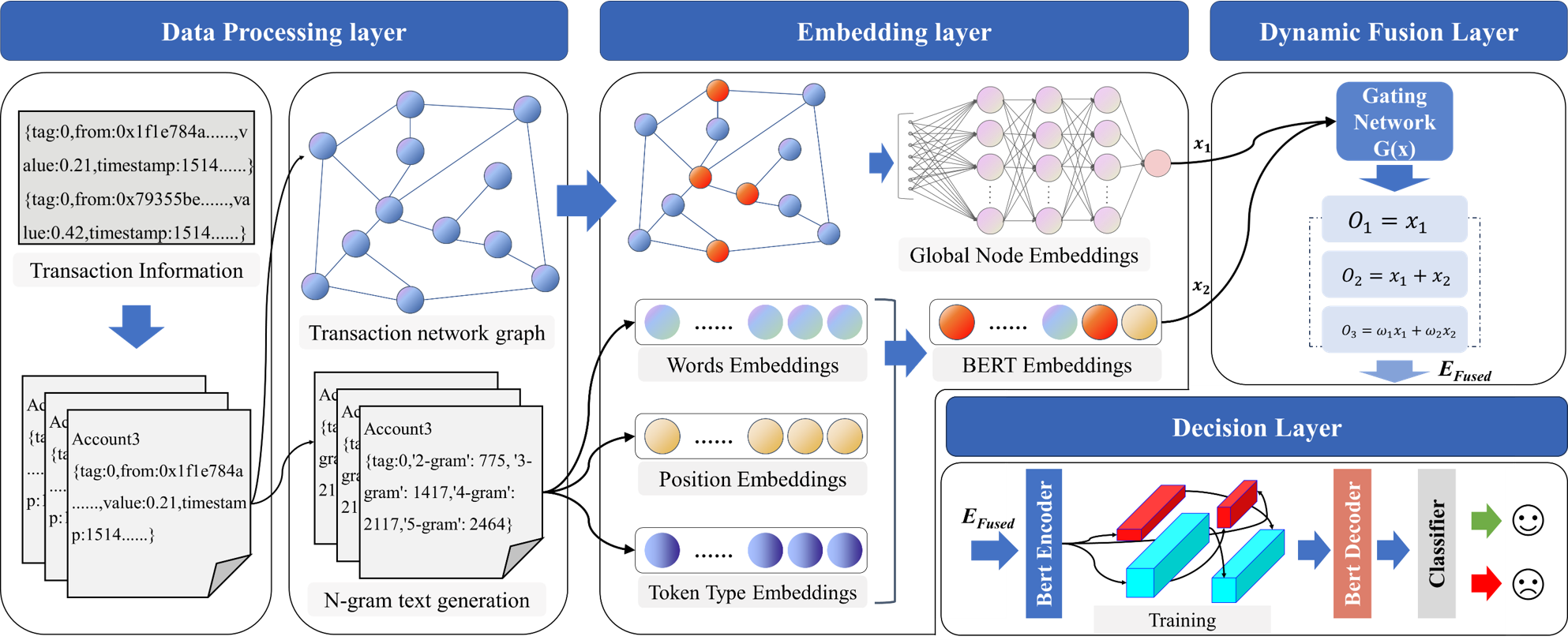}
    \caption{Architecture of the Dynamic Feature Fusion Model for Blockchain Fraud Detection}
    \label{1}
\end{figure*}

\section{Methodology}

In this chapter, we describe in detail a dynamic multimodal fusion approach for blockchain transaction data fraud detection. The proposed method integrates graph-based representation learning to capture global relationships within transaction networks and semantic feature extraction to identify local contextual patterns from transaction records. By leveraging a dynamic feature fusion mechanism, the model effectively combines structural and semantic information to enhance its ability to detect complex fraud behaviors, as illustrated in Figure \ref{1}.This chapter includes the detailed steps of our approach, starting with data generation and preprocessing, followed by a comprehensive explanation of the model architecture and the training process used to optimize performance.

\subsection{Data generation and pre-processing}
In the processing of blockchain transaction datasets, each transaction record typically contains several fields, such as \texttt{tag}, \texttt{from\_address} (sender address), \texttt{to\_address} (recipient address), \texttt{Value} (transaction Value), and \texttt{timestamp} (transaction timestamp). These fields describe the transaction behavior, the time it occurred, and the parties involved. To more effectively analyze and model transaction relationships, we need to properly classify and reorganize the transaction data.

Specifically, we classify all transaction data by sender and recipient addresses, constructing a transaction record structure based on accounts. This classification step not only simplifies transaction storage and access but also lays the foundation for subsequent graph structure construction and local semantic analysis.

Each transaction contains two account addresses, the sender (\texttt{from\_address}) and the recipient (\texttt{to\_address}). We classify transactions based on the sender's address (\texttt{from\_address}), treating it as the transaction record of an account. Each transaction is labeled as an "outgoing" transaction, with the field \texttt{in\_out} = 1. Similarly, when an account is the recipient, the transaction is labeled as an "incoming" transaction, with the field \texttt{in\_out} = 0. 

The classified transaction records are stored in a dictionary \texttt{accounts}, where the keys are account addresses and the values are lists of all transaction records for that account. Each list associated with an account contains all outgoing and incoming transactions related to that account. In this way, by separating and indexing transaction records by account, we can quickly retrieve the transaction history of any account, especially when analyzing account behavior patterns or transaction frequency.

\subsubsection{Time Aggregation Feature Enhancement}
To improve the information expression capability of transaction data in the time dimension, we particularly focus on the time aggregation characteristics of transactions during the data generation and preprocessing stages. By enhancing the time aggregation features, we can effectively capture some potential abnormal account behaviors, especially those accounts that engage in a large number of fund transactions within a short period \cite{chen2019exploiting}. These behaviors are often typical characteristics of phishing accounts, so analyzing and utilizing information in the time dimension is crucial for accurately detecting fraudulent activities.

When processing the transaction data of each account, we first sort the transaction records based on the timestamp. The purpose of sorting is to ensure that the subsequent time difference calculations reflect the actual order of the transactions, providing foundational support for time aggregation features. By sorting the transactions in chronological order, we can capture the flow of funds in an account over a specific period and further analyze the frequency and density of its transaction behavior.

To quantify the degree of frequent transactions in a short time, we introduce the n-gram time difference feature. Specifically, the n-gram time difference measures the compactness of transaction times by calculating the time difference between a transaction and the previous $n-1$ transactions. We calculate the time differences for 2-gram to 5-gram, with the formula as follows:
\[
\Delta T_{n} = T_{i} - T_{i-(n-1)}
\]
where $t_i$ denotes the timestamp of the $i$th transaction and $t_{i-(n-1)}$ denotes the timestamp of the $i-(n-1)$th transaction for the account. If the number of transactions is not sufficient to calculate the n-gram, the time difference is set to 0.

The n-gram time difference feature allows us to capture patterns of frequent trading over short periods. For example, if an account makes multiple inbound and outbound trades within a few minutes, the n-gram time difference will be significantly smaller, and this temporal aggregation reflects the account's high frequency of trades over a short period, which is often closely associated with phishing behavior.
\subsubsection{Graph Data Generation}

To effectively capture the inter-account relationships in blockchain transaction data, we first construct a graph-based data structure to represent the transaction network. In this section, we use an adjacency matrix \( \mathbf{A} \) to quantify the connection weights between accounts in the transaction network. The process of generating this graph representation involves the following steps:

\begin{enumerate}
    \item \textbf{Creating the Zero Matrix} \\
    We first create a \( n \times n \) zero matrix \( \mathbf{A} \), where \( n \) denotes the number of unique account addresses. This adjacency matrix is used to store the connection weights between different accounts. The elements of the matrix \( A[i,j] \) denote the transaction weights between account \( i \) and account \( j \).
    \[
    \mathbf{A} = \mathbf{0}_{n \times n}
    \]

    \item \textbf{Traversing Transaction Records} \\
    In order to populate the elements of the adjacency matrix, we need to iterate through all the transaction records \( T_k \), where each transaction \( T_k \) contains a sender \( \text{from\_address}_k \) and a receiver \( \text{to\_address}_k \). We use an `address\_to\_index` dictionary to map these account addresses to indices in the adjacency matrix. 

    \begin{itemize}
        \item The sender address is mapped as \( \text{from\_idx} \)
        \item The receiver address is mapped as \( \text{to\_idx} \)
    \end{itemize}

    The formulaic representation is as follows:

    \[
    \text{from\_idx} = \text{address\_to\_index}(\text{from\_address}_k)
    \]
    \[
    \text{to\_idx} = \text{address\_to\_index}(\text{to\_address}_k)
    \]

    \item \textbf{Calculating Transaction Weights} \\
    The weight of each transaction \( w_k \) reflects the transaction Value and the complexity of the transaction behavior. In order to capture more temporal features, we introduce a weight calculation method based on the n-gram time difference. The weight of a transaction \( w_k \) is obtained by weighted summation based on the n-gram time difference of the transaction. The formula for calculating the weights is as follows:

    \[
    w_k = \sum_{n=1}^{N} \alpha_n \cdot \Delta t_{n,k}
    \]

    Among them:
    \begin{itemize}
        \item \( \alpha_n \) denotes the weight coefficient for each n-gram time difference;
        \item \( \Delta t_{n,k} \) denotes the n-gram time difference of the transaction \( T_k \).
    \end{itemize}

    In addition, the transaction Value \( \text{Value}_k \) is also an important component of the weights, which we combine with the n-gram time difference to further adjust the weights of the transactions:

    \[
    w_k = \text{Value}_k \cdot \left( \sum_{n=1}^{N} \alpha_n \cdot \Delta t_{n,k} \right)
    \]

    \item \textbf{Populating the Adjacency Matrix} \\
    Once the weights of the transactions \( w_k \) are computed, we accumulate them to the corresponding positions in the adjacency matrix \( A[\text{from\_idx}, \text{to\_idx}] \). This accumulation operation allows the weights of multiple transactions to be superimposed, reflecting the strength of the connections between accounts:

    \[
    A[\text{from\_idx}, \text{to\_idx}] \mathrel{+}= w_k
    \]
\end{enumerate}

This operation ensures that when multiple transactions occur between two accounts, the corresponding weights are accumulated in the appropriate elements of the adjacency matrix. This accumulation process effectively reflects both the frequency of transactions and the aggregate transaction values between accounts. The resulting adjacency matrix \( \mathbf{A} \) serves as the input for graph-based representation learning, enabling the model to capture and analyze the global structural relationships within the transaction network.

\subsubsection{Text Transaction Data Generation}
In the transaction records of each account, the from\_address, to\_address and timestamp fields record the address information and timestamp of the account. Although these fields are important for transaction classification and temporal feature enhancement, they are not needed in text analysis, so we delete these fields before generating text data to simplify the data structure and retain key information such as transaction Value and label.
   
To prevent the model from relying too much on the sequential information of transactions in text processing, we randomly rearrange the transaction list for each account. This operation disrupts the backward and forward order of transactions, allowing the model to focus on the content features of transactions rather than time-dependent information, thus avoiding possible noise interference.

For example, the list of trades for account A is $[T_1,T_2,T_3]$ before disruption, and after random disruption may become$[T_2,T_1,T_3]. $

Next, we tag each account with an overall tag. An account is labelled as fraudulent whenever there is a transaction in the account with tag = 1, i.e., the account is labelled with a tag of 1. This tag is given to the first transaction record of the account. To simplify the transaction logging, the tag information for the rest of the transactions is deleted and only the tag of the first transaction is retained. This is because even if only one transaction in the account is related to fraud, the account itself may be potentially risky and may even be used for wider fraudulent activity. Typically, phishing accounts tend to mask their malicious behaviour by disguising multiple normal transactions. Therefore, in order to ensure the security and effectiveness of fraud detection, we have adopted more stringent criteria to ensure that the model can identify potentially high-risk accounts and prevent them from engaging in further illegal transactions. This labelling approach can help the model learn the risk characteristics of the accounts more accurately and improve the overall detection effectiveness.

When generating text data, we process the transaction records of each account and convert them into a single line of descriptive text. The key fields of each transaction (e.g., label tag, transaction value, etc.) are combined to create a compact textual representation that encapsulates the transaction information for the corresponding account. This step produces the raw text corpus, which serves as input for subsequent semantic feature extraction through a pre-trained text representation model.

For example, a transaction record might be converted into the following form:
$$Account A: tag= 1, Value= 5. 06854256, \ldots$$
The generated textual transaction dataset is partitioned in the ratio of 80\% training set, 10\% validation set, and 10\% test set. This data partitioning ensures that the model can learn enough features during the training process as well as perform performance tuning with the validation set, while verifying the model's generalisation ability on the test set.

\subsubsection{Text Data Cleaning}
After generating the textual transaction data, further pre-processing steps are applied to ensure compatibility with the input format required for the downstream semantic representation model. These steps include reading the generated TSV files, tokenizing the text into subword units, and transforming it into a format suitable for deep learning-based training.

We first read the generated Train.tsv and dev.tsv files, which contain the processed training set and validation set data. To ensure that the models are exposed to diverse data distributions during training, we randomly disrupt the data order to avoid overfitting the models to a specific data order. In addition, the test set data was read from test.tsv and similarly randomly disrupted.

After reading and shuffling the data, the training, validation, and test sets were combined into a unified data frame. From this, two key columns were extracted: the transaction text description (corpus) and the account label (y). The transaction text description captures the account's transaction behavior, while the label indicates whether the account is associated with fraudulent activity. This operation produces the input corpus and the corresponding supervisory signals (labels) required for subsequent semantic feature extraction and model training.

This was followed by a tokenization process to segment each document into a series of tokens (sub-word units), which were then normalized and encoded as necessary. This step ensures that the transaction text is transformed into a format suitable for semantic representation models, resulting in sequences of token IDs. These token IDs serve as inputs to the embedding layer of the text processing model for subsequent training. To ensure robustness, the order of documents is intentionally shuffled, exposing the model to unordered and varied inputs during training. Additionally, the labeled data y is aligned with the tokenized sentences and used as supervisory signals for the supervised learning process.

The dataset generated in the above steps contains global transactional relationships and local transactional semantic information, providing multimodal input for subsequent model training.

\subsection{ETH-GBERT Model Architecture}
To address the challenge of detecting complex fraudulent activities in blockchain transactions, we propose the ETH-GBERT Model, a deep learning framework designed to simultaneously capture global structural relationships and local semantic similarities. While transaction networks contain rich global patterns that reflect account interactions, transaction records hold local contextual details that can signal fraudulent behaviors. Existing methods often focus on one aspect, failing to leverage the complementary strengths of both.

In this study, we adopt Graph Convolutional Networks (GCNs) to capture the global transaction relationships embedded in account interaction graphs. GCNs are particularly suited for extracting structural features from graph-based data, making them ideal for modeling the relationships in blockchain transaction networks. Simultaneously, we use a pre-trained BERT model to analyze the local semantic features present in transaction text data, effectively capturing the contextual meaning and subtle patterns in transaction details.

By integrating these two components through a multimodal fusion mechanism, the ETH-GBERT Model combines insights from both global structural features and local semantic representations to enhance fraud detection performance. The following sections provide a detailed explanation of the architecture and design of the ETH-GBERT Model components.

\subsubsection{Model Architecture}
The ETH-GBERT Model integrates two core modules: GCN for dealing with trading account graphs and BERT for dealing with trading text data. Overall it can be divided into the following parts:

\begin{enumerate}
    \item \textbf{Graph-Based Representation Module}: Primarily captures global relationships within the transaction network. Through the GCN layers, the relationships between transaction accounts are convoluted, generating node embeddings (account embeddings) with global semantic information.
    \item \textbf{Semantic Feature Extraction Module}: Extracts local semantic information from transaction text data. The BERT model deeply represents the transaction records for each account and generates high-dimensional text embeddings.
    \item \textbf{Multimodal Fusion}: The GCN-generated global account embeddings and BERT-produced local text embeddings are fused, forming a multimodal embed vector. This fusion enables the model to take advantage of both the transaction network structure and the text features for fraud detection.
    \item \textbf{Classifier}: The fused embedding vector is passed through a fully connected layer for classification, outputting predictions to determine whether the account is related to fraudulent behavior.
\end{enumerate}

\subsubsection{Graph-Based Representation Module Design}

Adjacency Matrix Input. The input to the GCN module is the adjacency matrix \( \mathbf{A} \) of the transaction account graph, where the element \( A[i, j] \) represents the transaction weight between the account \( i \) and the account \( j \). This adjacency matrix is obtained from the graph data generation steps described earlier, incorporating transaction amounts and time features.

Graph Convolution Layer (GCN Layer). In the GCN module \cite{kipf2016semi}, the transaction account graph undergoes feature extraction through multiple graph convolution layers. The convolution operation in each layer is represented by the following formula:
\[
\mathbf{H}^{(l+1)} = \sigma\left( \tilde{\mathbf{D}}^{-\frac{1}{2}} \tilde{\mathbf{A}} \tilde{\mathbf{D}}^{-\frac{1}{2}} \mathbf{H}^{(l)} \mathbf{W}^{(l)} \right)
\]
where:
\begin{itemize}
    \item \( \mathbf{H}^{(l)} \) represents the node feature matrix at the \( l \)-th layer (account embedding matrix), and the initial \( \mathbf{H}^{(0)} \) is the initial feature of the transaction accounts;
    \item \( \tilde{\mathbf{A}} = \mathbf{A} + \mathbf{I} \) is the adjacency matrix with self-loops;
    \item \( \tilde{\mathbf{D}} \) is the degree matrix of the adjacency matrix;
    \item \( \mathbf{W}^{(l)} \) is the weight matrix at the \( l \)-th layer;
    \item \( \sigma \) is a non-linear activation function, such as ReLU.
\end{itemize}

Through multiple convolution operations, the model aggregates the global information of the transaction network layer by layer, eventually generating node embeddings with global transaction relationships.

\subsubsection{Semantic Feature Extraction Module Design}

Text Input and Initial Embeddings. The input to the BERT module is the transaction text data. After being cleaned and tokenized, the text data is converted into token sequences. These token sequences are embedded using BERT's Word Embedding, Position Embedding, and Token Type Embedding layers \cite{devlin2018bert}:

\[
\mathbf{E}_{\text{BERT}} = \mathbf{E}_{\text{word}} + \mathbf{E}_{\text{position}} + \mathbf{E}_{\text{token\_type}}
\]
where:
\begin{itemize}
    \item \( \mathbf{E}_{\text{word}} \) represents word embeddings;
    \item \( \mathbf{E}_{\text{position}} \) represents position embeddings to capture the positional relationships of words in the sequence;
    \item \( \mathbf{E}_{\text{token\_type}} \) represents token type embeddings to differentiate between different sentences or paragraphs.
\end{itemize}

BERT Encoding Layer. The embedded text is passed through BERT’s multi-layer Transformer encoder, generating higher-level text representations. BERT’s self-attention mechanism effectively captures the dependencies between different words in the transaction text, thus extracting local semantic similarity information.

The encoding operation of the Transformer \cite{vaswani2017attention} can be represented as follows:
\[
\mathbf{H}_{\text{fusion}} = \text{Transformer}(\mathbf{E}_{\text{Fused}})
\]
where
\begin{itemize}
    \item \( \mathbf{E}_{\text{Fused}} \) is calculated as presented in Section~\ref{multi}.
    \item \( \mathbf{H}_{\text{fusion}} \) is the text embedding generated by BERT.
\end{itemize}

\subsubsection{Multimodal Fusion}
\label{multi}
In the multimodal fusion stage of the model, we introduce a \textbf{dynamic feature fusion mechanism} inspired by DynMM~\cite{xue2023dynamic}, which adaptively determines the contributions of BERT and GCN embeddings for each input instance.

Fusion Strategy. Our approach employs a \textbf{gating network} $G(x)$ to generate instance-specific fusion weights. This allows the model to dynamically decide how much information to extract from BERT embeddings and BERT-enhanced GCN embeddings. Specifically, three fusion strategies are considered:
\begin{itemize}
    \item \textbf{BERT-only embeddings} $E_{\text{BERT}}$: Using textual information exclusively for prediction.
    \item \textbf{GCN-enhanced BERT embeddings} $E_{\text{GCN\_Enhanced}}$: GCN embeddings that integrate structural graph information and are enhanced with contextual features from BERT.
    \item \textbf{A weighted combination of BERT and GCN embeddings}: 
    \[
        E_{\text{Fusion}} = \alpha \cdot E_{\text{BERT}} + (1 - \alpha) \cdot E_{\text{GCN\_Enhanced}},
    \]
    where $\alpha$ is a learnable parameter initialized to $0.5$.
\end{itemize}

Dynamic Weight Calculation. The gating network $G(x)$ takes as input the concatenated features $[E_{\text{BERT}}, E_{\text{GCN\_Enhanced}}]$ and outputs fusion weights $g = [g_1, g_2, g_3]$ corresponding to the three fusion strategies:
\[
    g_i = \frac{\exp\left((\log G(x)_i + b_i)/\tau\right)}{\sum_{j=1}^3 \exp\left((\log G(x)_j + b_j)/\tau\right)}, \quad i \in \{1, 2, 3\},
\]
where $b \sim \text{Gumbel}(0, 1)$ is Gumbel noise, and $\tau$ is the temperature parameter controlling the smoothness of the output. When $\tau$ is large, the output weights $g_1, g_2, g_3$ are smooth and close to a soft probability distribution, enabling a soft fusion of the three strategies. As $\tau$ decreases, the output distribution becomes sharper, and when combined with the hard option, the weights are discretized into a one-hot vector through the straight-through estimator, effectively selecting a single fusion strategy while maintaining differentiability.

To handle varying task complexities and data characteristics, the gating network $G(x)$ can be implemented using different architectures, such as Multi-Layer Perceptrons (MLPs), Transformer layers, or convolutional networks.

In this work, we implement the gating network as a \textbf{Multi-Layer Perceptron (MLP)}, consisting of two fully connected layers with a ReLU activation function.

The final fused embedding $E_{\text{Fused}}$ is obtained as:
\[
    E_{\text{Fused}} = g_1 \cdot E_{\text{BERT}} + g_2 \cdot E_{\text{GCN\_Enhanced}} + g_3 \cdot E_{\text{Fusion}}.
\]

Adaptive Fusion Mechanism. This dynamic fusion mechanism enables the model to adapt its computational resources and fusion strategy based on the input complexity:
\begin{itemize}
    \item For \textbf{easy inputs}, the gating network assigns higher weights to simpler strategies such as $E_{\text{BERT}}$ or $E_{\text{GCN\_Enhanced}}$, reducing computational costs.
    \item For \textbf{complex inputs}, the gating network increases the contribution of the weighted combination $E_{\text{Fusion}}$, allowing the model to effectively integrate information from both modalities.
\end{itemize}

Compared to traditional static fusion approaches, our method fully and efficiently leverages the complementary strengths of BERT and GCN embeddings. By dynamically adjusting the fusion strategy for each input, the proposed approach achieves a better balance between \textbf{computational efficiency} and \textbf{representation power}.

\subsubsection{Classifier Design}
The fused multimodal embedding vector \( \mathbf{H}_{\text{fusion}} \) is input into a fully connected layer for the classification task. Through the Softmax layer, the model outputs the probability of whether an account is related to fraudulent behavior:
\[
\mathbf{y} = \text{Softmax}(\mathbf{W}_{\text{fusion}} \mathbf{H}_{\text{fusion}} + \mathbf{b}_{\text{fusion}})
\]
where \( \mathbf{W}_{\text{fusion}} \) and \( \mathbf{b}_{\text{fusion}} \) are the weight matrix and bias vector of the classifier, respectively.

ETH-GBERT Model enhances the joint learning of global relationships and local semantic information in blockchain transactions through the fusion of GCN and BERT embeddings. Through multimodal fusion, the model improves its ability to detect complex fraudulent behaviors effectively.

\section{Validation}
\subsection{Dataset review}
As shown in Table 1, we use three datasets as follows.

\subsubsection{Ethereum Phishing Transaction Network }This dataset is publicly available and is provided by Chen et al. (2021). The dataset is obtained by performing second-order breadth-first search (BFS) from known phishing nodes over a large-scale Ethernet transaction network. The dataset contains 2,973,489 nodes, 13,551,303 edges, and 1,165 phishing nodes\cite{chen2020phishing}.

\subsubsection{First-order Transaction Network of Phishing Nodes} This dataset was collected by Wu et al. (2022) through Ethernet nodes. It includes 1,259 phishing accounts and an equal number of normal accounts. The first-order neighbours of each account and the transaction edges between them are also included in the dataset, and the subnetwork contains about 60,000 nodes and 200,000 transaction edges\cite{wu2020phishers}.

\subsubsection{BERT4ETH} We use the BERT4ETH dataset provided by Hu et al. (2023), which is generated from sequences of Ether transactions and contains subsets of phishing accounts, de-anonymised data (ENS and Tornado Cash), and ERC-20 token logs.The BERT4ETH dataset is able to capture multihop relationships between trading accounts and is is suitable for phishing account detection and account de-anonymisation tasks. This dataset is an important component of our experiments and helps to further evaluate the performance of the model\cite{hu2023bert4eth}.

\subsection{Baseline}
In this experiment, we selected three common categories of baseline models for comparison:
\begin{enumerate}
    \item Graph embedding methods based on random walks, including DeepWalk~\cite{perozzi2014deepwalk}, Trans2Vec~\cite{wu2020phishers}, Dif2Vec~\cite{rozemberczki2018fast}, and Role2Vec~\cite{ahmed2018learning,beres2021blockchain};
    \item Graph neural network(GNN) models, including GCN~\cite{kipf2016semi}, GraphSAGE~\cite{hamilton2017inductive}, and GAT~\cite{velivckovic2017graph};
    \item BERT4ETH, a model designed specifically for fraud detection on Ethereum~\cite{hu2023bert4eth}.
\end{enumerate}

DeepWalk generates node sequences through random walks on the graph and employs the skip-gram model to learn low-dimensional representations of nodes. Trans2Vec builds on DeepWalk by incorporating transaction heterogeneity and temporal features, designed specifically for detecting phishing accounts in the Ethereum network. Dif2Vec adjusts the sampling probabilities of nodes during random walks to enhance the diversity of embeddings by increasing the sampling of low-degree nodes. Role2Vec learns structural roles of nodes rather than focusing solely on proximity relationships, generating more generalizable embeddings. 

Regarding GNN-based models, GCN aggregates the features of neighboring nodes via convolution operations to learn node representations, making it suitable for tasks such as node classification. GraphSAGE generates new node embeddings by sampling and aggregating the features of neighboring nodes, which enables it to handle large-scale graph data. GAT introduces an attention mechanism, dynamically assigning weights to each node's neighbors to aggregate node information more effectively. 

BERT4ETH is specifically designed for detecting fraudulent activities on the Ethereum network, leveraging BERT along with transaction data features from the Ethereum network to identify fraudulent behavior within blockchain transactions.

In our experiments, all baseline models, including BERT4ETH, DeepWalk, Trans2Vec, Dif2Vec, Role2Vec, GCN, GSAGE, and GAT, were implemented according to the original configurations specified in their respective papers. This ensures a fair comparison of performance across different models.

\section{Preprocessing and Training Settings}

In this section, we describe the ETH-GBERT preprocessing setup, initial parameters, loss function, and evaluation metrics used in our experiment.

\subsection{Data Loading and Preprocessing}
Before training, the dataset was split into training, validation, and test sets, accounting for 80\%, 10\%, and 10\% of the total data, respectively. We used PyTorch's \texttt{DataLoader} to load the data in mini-batches, with shuffling applied during the training process. The training set is used to update model parameters, the validation set evaluates the model's generalization ability, and the test set is used for final performance evaluation.

\subsection{Hyperparameter Settings}
The following hyperparameters were set during the model training:
\begin{itemize}
    \item \textbf{Learning rate}: The initial learning rate was set to \(8 \times 10^{-6}\), and a learning rate scheduler was employed to adjust the learning rate dynamically.
    \item \textbf{Regularization coefficient}: L2 regularization was applied with a coefficient of \( \lambda = 0.001 \) to prevent overfitting.
    \item \textbf{Batch size and gradient accumulation}: The batch size was set to 32. We adopted gradient accumulation to save memory, updating the model's parameters after every 2 mini-batches.
    \item \textbf{Epochs}: The total number of training epochs was set to 40 for initial validation of the model's effectiveness.
\end{itemize}

\subsection{Loss Function and Optimizer}
We used the cross-entropy loss function for the classification task \cite{lecun2015deep}, defined as:
\[
\mathcal{L} = - \frac{1}{N} \sum_{i=1}^{N} \left( y_i \log(p_i) + (1 - y_i) \log(1 - p_i) \right)
\]
where \(N\) is the batch size, \(y_i\) is the ground truth label, and \(p_i\) is the predicted probability.

The AdamW optimizer was employed for optimization, combining the adaptive learning rate of Adam with L2 regularization through weight decay. The update rule for AdamW is given by:
\[
\theta_{t+1} = \theta_t - \eta \cdot \frac{m_t}{\sqrt{v_t} + \epsilon}
\]
where \(m_t\) and \(v_t\) are the first and second moments of the gradients, and \( \epsilon \) is a small constant to avoid division by zero.

\subsection{Evaluation Metrics}
At the end of each epoch, the model's performance was evaluated on the validation set using precision, recall, and F1 score as evaluation metrics:
\begin{itemize}
    \item \textbf{Precision}:
    \[
    \text{Precision} = \frac{\text{TP}}{\text{TP} + \text{FP}}
    \]
    \item \textbf{Recall}:
    \[
    \text{Recall} = \frac{\text{TP}}{\text{TP} + \text{FN}}
    \]
    \item \textbf{F1 Score}:
    \[
    \text{F1 Score} = 2 \cdot \frac{\text{Precision} \cdot \text{Recall}}{\text{Precision} + \text{Recall}}
    \]
\end{itemize}
Here, TP, TN, FP, and FN represent the number of true positives, true negatives, false positives, and false negatives, respectively. These evaluation metrics provide a comprehensive view of the model's classification performance and help monitor the generalization ability throughout the training process.

\begin{table*}[!ht]
\centering
\caption{Performance Comparison of ETH-GBERT and Baseline Models on Various Datasets}
\label{tab:results}
\begin{adjustbox}{width=\textwidth} 
\begin{tabular}{lccc ccc ccc}
\toprule
\textbf{Model} & \multicolumn{3}{c}{\textbf{Multigraph}} & \multicolumn{3}{c}{\textbf{Transaction Network}} & \multicolumn{3}{c}{\textbf{B4E}} \\ 
 & \textbf{F1 Score} & \textbf{Recall} & \textbf{Precision} & \textbf{F1 Score} & \textbf{Recall} & \textbf{Precision} & \textbf{F1 Score} & \textbf{Recall} & \textbf{Precision} \\ 
\midrule
BERT4ETH   & 67.11 & 61.25 & 74.21 & 64.21 & 62.17 & 66.39 & 64.26 & 63.58 & 64.95 \\ 
DeepWalk   & 58.44 & 58.21 & 58.67 & 59.21 & 58.31 & 60.14 & 54.51 & 55.38 & 53.67 \\ 
Trans2Vec  & 52.13 & 51.36 & 52.92 & 54.28 & 56.26 & 52.43 & 55.31 & 54.96 & 55.66 \\ 
Dif2Vec    & 65.27 & 64.21 & 66.37 & 62.11 & 62.54 & 61.69 & 63.25 & 63.54 & 62.96 \\ 
Role2Vec   & 74.13 & 74.52 & 73.74 & 71.39 & 71.58 & 71.20 & 74.25 & 74.25 & 74.25 \\ 
GCN        & 42.29 & 74.07 & 29.59 & 41.12 & 73.37 & 28.56 & 64.71 & 72.68 & 58.31 \\ 
GSAGE      & 35.47 & 34.77 & 36.20 & 33.79 & 32.99 & 34.64 & 53.28 & 60.47 & 47.62 \\ 
GAT        & 39.98 & 79.82 & 26.67 & 41.61 & 77.56 & 28.43 & 61.50 & 85.20 & 48.12 \\ 
\rowcolor{gray!20} \textbf{ETH-GBERT}  & \textbf{94.71} & \textbf{94.71} & \textbf{94.71} & \textbf{86.16} & \textbf{87.82} & \textbf{84.56} & \textbf{89.79} & \textbf{89.57} & \textbf{90.84} \\ 
\bottomrule
\end{tabular}
\end{adjustbox}
\end{table*}

\section{Performance}

To evaluate the effectiveness of our proposed ETH-GBERT model in detecting fraud in blockchain transaction data, we compared its performance with several baseline models, including BERT4ETH, DeepWalk, Trans2Vec, Dif2Vec, Role2Vec, GCN, GSAGE, and GAT. These models were applied to three different datasets: Multigraph, Transaction Network, and B4E. The comparison focuses on key metrics such as F1 score, recall, and precision, as shown in Table~\ref{tab:results}.

\begin{figure*}[!ht]
    \centering
    \includegraphics[width=\linewidth]{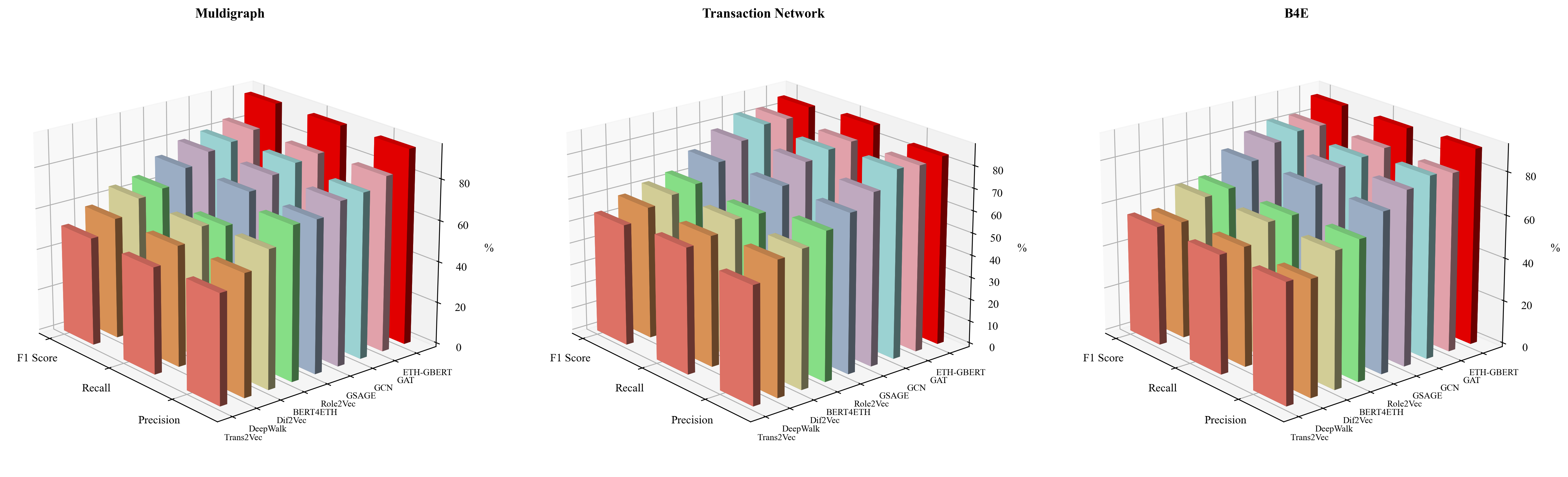}
    \caption{3D Comparative Analysis of Model Performance Metrics Across Datasets}
    \label{2}
\end{figure*}

\begin{figure}[!ht]
    \centering
    \includegraphics[width=\columnwidth]{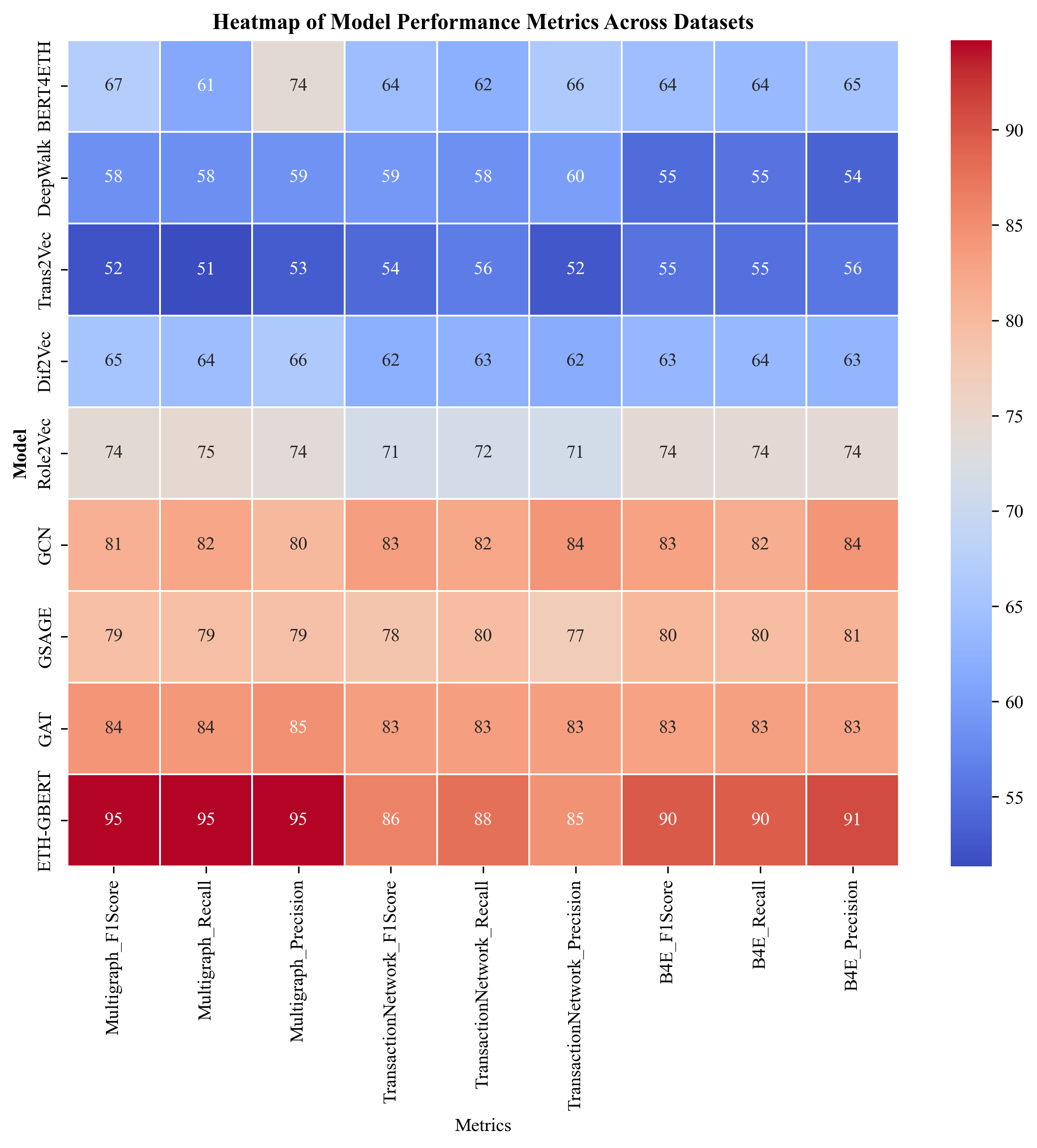}
    \caption{Heatmap of Model Performance Metrics Across Multiple Datasets}
    \label{3}
\end{figure}

\subsection{Overview of Model Performance}
From the experimental results, it is evident that ETH-GBERT significantly outperforms all baseline models across the datasets in terms of F1 score, recall, and precision. This is visually illustrated in Figure \ref{2}, which shows a 3D comparative analysis of model performance metrics across multiple datasets. Additionally, Figure \ref{3} provides a heatmap representation of model performance metrics, offering a detailed comparison across different models and metrics.
\begin{itemize}
    \item \textbf{On the Multigraph dataset}, ETH-GBERT achieves an F1 score of 94.71, approximately 10 points higher than GAT (84.35). This demonstrates ETH-GBERT's ability to effectively combine graph structures and semantic information for superior fraud detection performance.
    
    \item \textbf{On the Transaction Network dataset}, ETH-GBERT achieves an F1 score of 86.16, with a recall of 87.82 and precision of 84.56. Compared to GAT (F1 score of 83.27) and GCN (F1 score of 83.29), ETH-GBERT provides enhanced accuracy in capturing the complexities of transaction relationships.
    
    \item \textbf{On the B4E dataset}, ETH-GBERT achieves an F1 score of 89.79, surpassing all the baseline models. Notably, ETH-GBERT excels in recall, achieving 89.57, highlighting its sensitivity in identifying potential fraud cases.
\end{itemize}

\subsection{Comparison with Baseline Models}
Several key insights can be drawn from the comparison with baseline models:

\begin{enumerate}
    \item \textbf{BERT4ETH}: While BERT4ETH demonstrates reasonable performance in extracting local semantic information, its F1 scores on both the Multigraph and Transaction Network datasets (67.11 and 64.21, respectively) are significantly lower than ETH-GBERT. This highlights the importance of incorporating global structure information, which BERT4ETH lacks.

    \item \textbf{GCN and GSAGE}: GCN and GSAGE struggle to achieve competitive F1 scores, with GCN scoring 42.29 on the Multigraph dataset and 41.12 on the Transaction Network dataset. These models are effective in capturing global transaction relationships but lack the ability to integrate local semantic information, limiting their performance in fraud detection tasks.

    \item \textbf{GAT}: The GAT model benefits from its self-attention mechanism, achieving a relatively higher recall (e.g., 79.82 on the Multigraph dataset). However, its F1 scores remain low (39.98 on Multigraph and 41.61 on Transaction Network), due to its limited ability to model textual features and complex fraud patterns.

    \item \textbf{ETH-GBERT}: Our proposed ETH-GBERT model significantly outperforms all baseline models across all datasets. It achieves the highest F1 scores of 94.71, 86.16, and 89.79 on the Multigraph, Transaction Network, and B4E datasets, respectively. This performance demonstrates the effectiveness of ETH-GBERT in dynamically fusing global transaction network information with local semantic features from transaction texts, enabling superior fraud detection capabilities.
\end{enumerate}

\begin{table*}[!ht]
\centering
\caption{Performance Improvement Analysis via Multimodal Dynamic Fusion}
\label{tab:Dyn}
\begin{adjustbox}{width=\textwidth} 
\begin{tabular}{lcccccccccc}
\toprule
\textbf{Model} & \multicolumn{3}{c}{\textbf{Multigraph}} & \multicolumn{3}{c}{\textbf{Transaction Network}} & \multicolumn{3}{c}{\textbf{B4E}} \\ 
 & \textbf{F1 Score} & \textbf{Recall} & \textbf{Precision} & \textbf{F1 Score} & \textbf{Recall} & \textbf{Precision} & \textbf{F1 Score} & \textbf{Recall} & \textbf{Precision} \\ 
\midrule
BERT Only & 90.10 & 90.07 & 90.15 & 78.77 & 78.12 & 83.82 & 83.58 & 83.05 & 87.44 \\ 
\textit{Difference(\%)} & \textit{-4.61} & \textit{-4.64} & \textit{-4.56} & \textit{-7.39} & \textit{-9.70} & \textit{-0.74} & \textit{-6.21} & \textit{-6.52} & \textit{-3.40} \\ 
\hdashline
GCN Only  & 42.29 & 74.07 & 29.59 & 41.12 & 73.37 & 28.56 & 64.71 & 72.68 & 58.31 \\ 
\textit{Difference(\%)} & \textit{-52.42} & \textit{-20.64} & \textit{-65.12} & \textit{-45.04} & \textit{-14.45} & \textit{-56.00} & \textit{-25.08} & \textit{-16.89} & \textit{-32.53} \\ 
\hdashline
Simple Combination & 84.55 & 84.15 & 86.29 & 83.27 & 83.75 & 83.55 & 85.35 & 88.16 & 82.71 \\ 
\textit{Difference(\%)} & \textit{-10.16} & \textit{-10.56} & \textit{-8.42} & \textit{-2.89} & \textit{-4.07} & \textit{-1.01} & \textit{-4.44} & \textit{-1.41} & \textit{-8.13} \\ 
\hdashline
Weighted Combination & 92.43 & 92.51 & 92.47 & 84.18 & 83.75 & \textbf{86.73} & 86.76 & 86.34 & 90.20 \\ 
\textit{Difference(\%)} & \textit{-2.28} & \textit{-2.20} & \textit{-2.24} & \textit{-1.98} & \textit{-4.07} & \textit{+2.17} & \textit{-3.03} & \textit{-3.23} & \textit{-0.64} \\ 
\hdashline
\rowcolor{gray!20} \textbf{ETH\_GBERT} & \textbf{94.71} & \textbf{94.71} & \textbf{94.71} & \textbf{86.16} & \textbf{87.82} & 84.56 & \textbf{89.79} & \textbf{89.57} & \textbf{90.84} \\ 
\bottomrule
\end{tabular}
\end{adjustbox}
\end{table*}

\begin{table*}[!ht]
\centering
\caption{Performance with Different Normal to Fraud Ratios}
\label{tab:ratio_results}
\begin{adjustbox}{width=\textwidth} 
\begin{tabular}{lccc ccc ccc}
\toprule
\textbf{Ratio} & \multicolumn{3}{c}{\textbf{Multigraph}} & \multicolumn{3}{c}{\textbf{Transaction Network}} & \multicolumn{3}{c}{\textbf{B4E}} \\ 
 & \textbf{F1 Score} & \textbf{Recall} & \textbf{Precision} & \textbf{F1 Score} & \textbf{Recall} & \textbf{Precision} & \textbf{F1 Score} & \textbf{Recall} & \textbf{Precision} \\ 
\midrule
1:9   & 78.50 & 80.10 & 77.90 & 75.20 & 76.90 & 74.50 & 70.30 & 72.20 & 69.80 \\ 
2:8   & 81.30 & 82.40 & 80.20 & 77.80 & 79.50 & 76.70 & 73.10 & 74.80 & 72.90 \\ 
3:7   & 83.70 & 84.50 & 82.90 & 80.30 & 81.80 & 79.90 & 75.40 & 77.20 & 74.60 \\ 
4:6   & 87.50 & 88.20 & 86.70 & 84.10 & 85.40 & 83.82 & 79.10 & 80.70 & 78.90 \\ 
5:5   & \textbf{94.71} & \textbf{94.71} & \textbf{94.71} & \textbf{86.16} & \textbf{87.82} & \textbf{84.56} & \textbf{89.79} & \textbf{89.57} & \textbf{90.84} \\ 
6:4   & 89.30 & 90.20 & 88.70 & 83.80 & 85.10 & 82.90 & 80.10 & 82.60 & 79.80 \\ 
7:3   & 85.60 & 86.50 & 84.80 & 80.90 & 82.30 & 79.10 & 77.20 & 78.80 & 76.50 \\ 
8:2   & 82.30 & 83.40 & 81.60 & 77.60 & 79.10 & 76.40 & 73.90 & 75.50 & 73.20 \\ 
9:1   & 80.10 & 81.20 & 79.30 & 75.40 & 76.83 & 74.60 & 71.30 & 72.80 & 70.50 \\ 
\bottomrule
\end{tabular}
\end{adjustbox}
\end{table*}

\subsection{Improvement Analysis via Multimodal Dynamic Fusion}
The Table~\ref{tab:Dyn}  illustrates the performance improvements brought by multimodal dynamic fusion, showcasing the different performances of \textbf{Unimodal Models}, \textbf{Static Fusion Methods}, and \textbf{Dynamic Fusion}.

\begin{itemize}

    \item\textbf{Unimodal Models:} The BERT-only model achieves strong results in the Multigraph dataset (F1 Score = 90.10) due to its ability to model language-centric features. However, it performs poorly in Transaction Network and B4E datasets (F1 Scores = 78.77 and 83.58), indicating its limitations in graph-based tasks. Conversely, the GCN model, which relies solely on graph information, performs poorly across all datasets, showing its limited capacity to model textual features.

    \item\textbf{Static Fusion Methods:} The GCN-enhanced BERT approach combines semantic and graph features, but its fixed fusion mechanism prevents dynamic weight adjustment to fully leverage their respective advantages. As a result, the performance gains are limited, and on some datasets, the metrics even perform worse compared to using BERT alone.

    \item\textbf{Dynamic Fusion (ETH-GBERT):} The ETH-GBERT model, leveraging dynamic fusion, achieves the best performance across almost all datasets. It dynamically adjusts the contributions of BERT and GCN, resulting in F1 Scores of 94.71, 86.16, and 89.79 in Multigraph, Transaction Network, and B4E datasets, respectively. Compared to static fusion, it offers consistent improvements (e.g., +2.28 in Multigraph, +3.03 in B4E).
\end{itemize}

The results highlight the limitations of unimodal and static fusion methods in handling multimodal data. Dynamic fusion, as implemented in ETH-GBERT, effectively balances textual and graph-based features, achieving superior performance and adaptability across diverse tasks. This demonstrates its potential as a robust multimodal learning approach.

\subsection{Impact of Normal to Fraud Ratio on Model Performance}
In this subsection, we evaluate how varying the ratio of normal to fraud transactions in the dataset affects the performance of the ETH-GBERT model across three different datasets: Multigraph, Transaction Network, and B4E.

We trained the ETH-GBERT model on datasets with varying ratios of normal to fraud transactions, ranging from 1:9 to 9:1. The key evaluation metrics—F1 Score, Recall, and Precision—were tracked for each dataset to understand the impact of different data distributions on the model's performance.

\begin{figure}[!ht]
    \centering
    \includegraphics[width=\columnwidth]{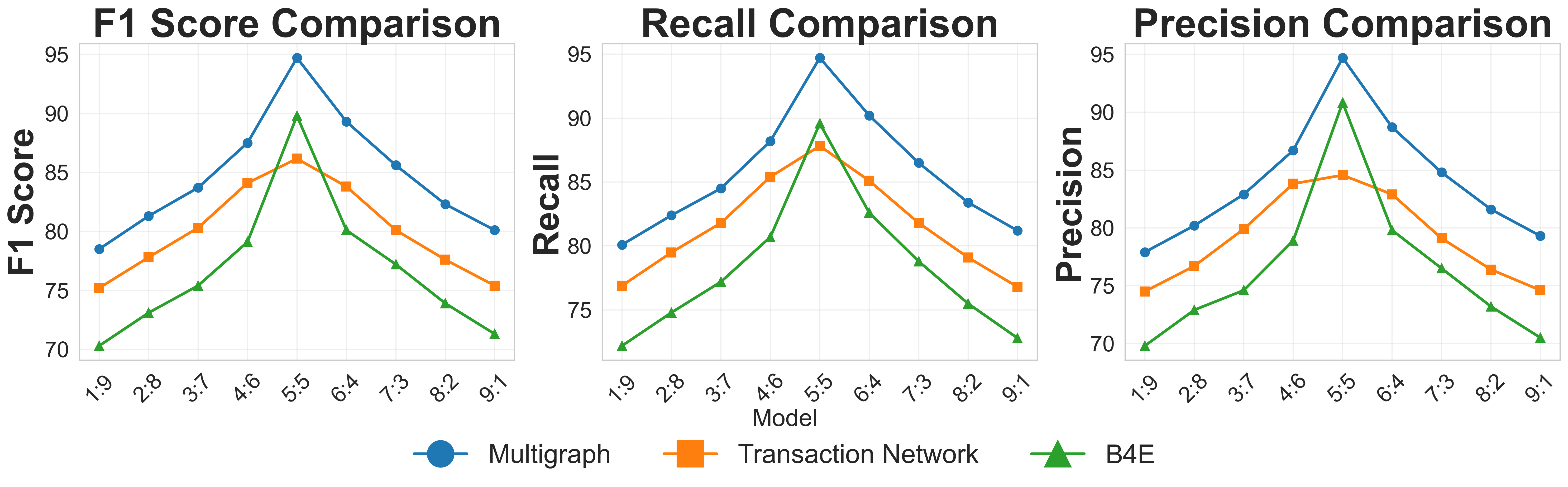}
    \caption{Trends of F1 Score, Recall, and Precision Across Different Normal-to-Fraud Ratios}
    \label{4}
\end{figure}

Table~\ref{tab:ratio_results} presents the performance metrics for each dataset under different normal-to-fraud ratios. The results demonstrate that the ETH-GBERT model performs optimally on a balanced dataset (5:5 ratio), achieving the highest F1 Score, Recall, and Precision. For example, in the \textbf{Multigraph} dataset, the model reaches an F1 Score of 94.71, while in the \textbf{Transaction Network} dataset, the highest F1 Score is 86.16. The \textbf{B4E} dataset also shows strong performance, with an F1 Score of 89.79 at the 5:5 ratio.

Although the performance declines as the data becomes imbalanced, the ETH-GBERT model remains robust. As illustrated in the figure \ref{4}, the overall performance decrease is more pronounced in datasets with more complex transaction patterns, such as the \textbf{B4E} dataset, where the interaction between normal and fraud transactions may contain more nuanced features.

These findings suggest that while the ETH-GBERT model can handle imbalanced datasets, a balanced ratio between normal and fraud transactions helps the model achieve its best performance.

\section{Conclusion}

In this paper, we proposed a novel dynamic multimodal fusion model(ETH-GBERT) for fraud detection in blockchain transactions. By adaptively integrating global structural features from transaction networks and local semantic information from transaction texts, the model effectively addresses the limitations of existing methods, achieving a better balance between computational efficiency and representation learning power.

To support the proposed model, we developed a comprehensive data processing pipeline, including graph construction for capturing inter-account relationships and temporal feature extraction using n-gram time differences. This pipeline enables the model to simultaneously analyze global structural patterns and local contextual features embedded within transaction data. Furthermore, the dynamic fusion mechanism introduced in this work adaptively adjusts the contributions of structural and semantic features based on transaction context, enhancing the model's accuracy and robustness in detecting complex fraudulent activities.

Through extensive experiments on large-scale blockchain datasets, our model demonstrated significant improvements over existing benchmark methods, achieving the highest F1 scores across multiple evaluation scenarios.

The key contributions of this study are as follows:
\begin{itemize} \item Proposing a multimodal fusion framework that dynamically integrates structural and semantic information to enhance blockchain fraud detection.
\item Developing a robust and efficient data processing pipeline that captures both global transaction relationships and temporal behavioral patterns.
\item Introducing a dynamic feature fusion mechanism that adaptively balances feature contributions, improving detection precision and efficiency across varied contexts.
\item Demonstrating the effectiveness of the proposed approach through experiments, where it significantly outperformed state-of-the-art models on multiple real-world datasets.
\end{itemize}

\bibliographystyle{elsarticle-num} 
\bibliography{my_ref}

\end{document}